\mag=\magstephalf
\pageno=1
\input amstex
\documentstyle{amsppt}
\TagsOnRight
\interlinepenalty=1000
\NoRunningHeads

\pagewidth{16.0 truecm}
\pageheight{23.0 truecm}
\vcorrection{-0.5cm}
\nologo

\font\twobf=cmbx12

\define \CC{{\Bbb C}}

\define \ZZ{{\Bbb Z}}
\define \NN{{\Bbb N}}

\define \tvskip{\vskip 1.0 cm}
\define\ce#1{\lceil#1\rceil}
\define\dg#1{(d^{\circ}\geq#1)}
\define\Dg#1#2{(d^{\circ}(#1)\geq#2)}

\define\s#1{\sigma_{#1}}
\define\tp#1{\negthinspace\left.\ ^t#1\right.}
\define\mrm#1{\text{\rm#1}}
\define\lr#1{^{\sssize\left(#1\right)}}

{\centerline{\bf{ Elliptic and Hyperelliptic Solutions of
Discrete Painlev\'e I }}}

{\centerline{\bf{ and Its Extensions to
Higher Order Difference Equations}}}

\author
Shigeki MATSUTANI${}^0$
\endauthor
\affil
8-21-1 Higashi-Linkan Sagamihara 228-0811 Japan
\endaffil
 \endtopmatter

\footnotetext{e-mail:RXB01142\@nifty.ne.jp}

\vskip 0.5 cm
\centerline{\twobf Abstract }\vskip 0.5 cm

The solutions of the discrete Painlev\'e equation I
 were constructed in terms of
elliptic and hyperelliptic $\psi$ functions for
algebraic curves of genera one and two.
For the case of genus two, there appear higher order
difference equations which naturally contain
 the discrete Painlev\'e equation I as a special case.

\vskip 0.5 cm
\centerline{\twobf Keywords }\vskip 0.5 cm

Discrete Painlev\'e Equation, Elliptic $\psi$ Function, Hyperelliptic
$\psi$ Function


\vskip 0.5 cm
\centerline{\twobf \S 1. Introduction }\vskip 0.5 cm

In this article, we will consider the algebraic solutions of
the discrete Painlev\'e I equation [ORGT]
$$
        \beta_{n+1}\beta_{n-1} = \frac{z}{\beta_{n}}+
 \frac{a}{\beta_{n}^2}, \tag 1-1
$$
where $z$ and $a$ are some parameters.
We give solutions of the equation (1-1)
in terms of the elliptic and hyperelliptic $\psi$ functions.

In section 2, we give an elliptic solution of this equation.
Section 3 contains our main subject, which is
based upon the recent studies on $\psi$
function [C, Ma2, M\^O, \^O1, \^O2].
The $\psi$ function is defended over an algebraic curve itself
embedded in its Jacobian rather than over the Jacobian variety.
Although they need slightly corrections, Cantor essentially
gave a determinant expression of $\psi$-function and
a recursion equation on the $\psi$-functions
of genus two [C]. On the other hand, \^Onishi gave
another determinant expression of $\psi$-function [\^O2].
Recently both expressions are connected by us [M\^O].
In \S3, we show that the recursion relation with a
correction become (1-1)
and a natural third order difference equation
for a certain point in the related algebraic curve,
$$
        d_{m+2}d_{m-1} =\frac{ \alpha_5}{d_{m+1} d_m}
                - \alpha_4
          \left( \frac{1}{d_{m+1}}+\frac{1}{d_m}\right),
 \tag 1-2
$$
where $\alpha$'s are some parameters. (1-2) should
be regarded as an extension of (1-1).
By using our identification [M\^O] and the expressions
of \^Onishi, we give their solutions in terms of the
hyperelliptic $\sigma$ functions [B1-3].
We also obtain the sixth and forth order ordinary difference
equations related to genus two curves.

This study started after listing to the lecture
of the recent progress of third order
difference equation by
 Yahagi, Kimura, Tusjimoto and Hirota [YKTH, HYK].
They have been studying the
third order difference equations.
Their motivation is to construct a list of the
integrable ordinary third order
differential equations as Painlev\'e school gave one for
second order one
in beginning of the last century. Recent studies of
difference equations show that integrablity, at least,
for the case
of the ordinary second differential equations is determined by
properties of their difference equations versions. Hence
in order to construct the list, they started to classify
the third order difference equations
by aided of numerical computations.
They have a list which contains eleven types of
 the third order difference equations.
They stated that they found an equation whose solution
can not be expressed by any elliptic functions.
As I believe that some class of the difference equations
must be defined over an algebraic curve itself embedded in
the Jacobi variety whereas
the  continuous soliton equation
is related to theory of Jacobi variety rather than curve
itself.
Thus I began to study this problem along the line of
arguments of \^Onishi [\^O1,2] and Cantor [C].

\vskip 0.5 cm
\centerline{\twobf \S 2. Elliptic $\psi$-Function }\vskip 0.5 cm

In this section, we consider an elliptic curve,
$$
   y^2 = 4 x^{3} - g_2 x -g_3, \tag 2-1
 $$
where $g$'s are complex numbers.

The elliptic $\psi$-function \cite{W} is defined as
$$
        \psi_n(u) = \frac{\sigma(nu)}{\sigma(u)^{n^2}},
\tag 2-2
$$
where $\sigma(u)$ is the Weierstrass sigma function.
This function has  Brioschi-Kiepert
formula \cite{W-W, p.460 and refecences in \^O2},
$$
 \psi(u)= (-1)^{n(n-1)/2}(1!2!\cdots n!)^2
   \frac{\sigma(nu)}{\sigma(u)^{n^2}}
  = \left|\matrix
    \wp'(u)        & \wp''(u)     &  \cdots  & \wp\lr{n-1}(u)  \\
    \wp''(u)       & \wp'''(u)    &  \cdots  & \wp\lr{n}(u)    \\
    \vdots         & \vdots       &  \ddots  & \vdots          \\
    \wp\lr{n-1}(u) & \wp\lr{n}(u) &  \cdots  & \wp\lr{2n-3}(u) \\
   \endmatrix\right|.
  \tag 2-3
$$

The $\psi$ function also obeys the recursion relation,
$$
 \psi_{n+m} \psi_{m-n}
=\left| \matrix
 \psi_{m-1}\psi_n & \psi_{m}\psi_{n+1} \\
\psi_{m}\psi_{n-1} & \psi_{m+1}\psi_{n}
\endmatrix \right|. \tag 2-4
$$
This relation is proved by the additive formula of
Weierstrass $\wp$ function [T]. We note that this recursion relation
differs from the identities of Hankel determinant (2-3).
For $m=2$ case, we have a bilinear difference equation,
$$
\psi_{n+2}\psi_{n-2} -
        \psi_2^2\psi_{n-1}\psi_{n+1}
+\psi_3\psi_1\psi_{n}\psi_{n} =0. \tag 2-5
$$
By introducing the quantity $\beta_n = \psi_{n+1}
\psi_{n-1}/\psi_{n}^2$,
(2-5) turns out to be the difference equation,
$$
        \beta_{n+1}\beta_{n-1} = \frac{\psi_2^2}{\beta_{n}}-
 \frac{\psi_3\psi_1}{\beta_{n}^2}. \tag 2-6
$$
This equation is identified with the discrete Painlev\'e I (1-1)
\cite{ORGT}.
We have a special solution of (1-1) as
$$
        \beta_n = \frac{\sigma((n+1)u) \sigma((n-1)u)}
        {\sigma(u)^2\sigma(ru)^2}.\tag 2-7
$$

\vskip 0.5 cm
\centerline{\twobf \S 3. Hyperelliptic $\psi$-function of Genus Two}
\vskip 0.5 cm

In this section, we will deal with the $\psi$-function over
 a hyperelliptic curve $C$  of genus two defined by
an affine equations,
$$
 \split
   y^2 &= f(x) \\
   &= \lambda_0 +\lambda_1 x
        +\lambda_2 x^2  +\lambda_3 x^3 +\lambda_4 x^4
        +\lambda_5 x^{5},
\endsplit \tag 3-1
 $$
where $\lambda_{5}\equiv1$ and $\lambda_j$'s are complex numbers.
We denote its corresponding Jacobi variety by $\Cal J$ and image
of inclusion of the curve into $\Cal J$ by $\iota(C)$.

The hyperelliptic $\psi$ function of genus two is given by
[\^O1, \^O2],
$$
        \psi_n(u)=\frac{\sigma( nu )}{\sigma_2( u)^{n^2}}.
          \tag 3-2
$$
Here $u:=(u_1,u_2)$ is a coordinate restricted to curve
itself $\iota (C)$,
$$
        u_1:=\int^{(x,y)}_\infty \frac{dx}{y},\quad
        u_2:=\int^{(x,y)}_\infty \frac{xdx}{y},\tag 3-3
$$
in the Jacobi variety $\Cal J:=\{(\tilde u_1, \tilde u_2)\}$,
$$
        \tilde u_1:=\int^{(x,y)}_\infty \frac{dx}{y}+
          \int^{(x_2,y_2)}_\infty \frac{dx}{y},\quad
        \tilde u_2:=\int^{(x,y)}_\infty \frac{xdx}{y}+
          \int^{(x_2,y_2)}_\infty \frac{xdx}{y},\tag 3-4
$$
for $(x,y), (x_2,y_2) \in C$.
Further $\sigma$ is Baker's sigma function [Ba1, Ba2, Ba3]
and $\sigma_2$ is its derivative with respect to $u_2$.
In this article, we will follow the arguments in \cite{\^O1, \^O2, M\^O}.
We note that $u_1$ is a function of $u_2$ \cite{M\^O}.
This $\psi_n(x,y)$ is a polynomial whose
zero $(x_0,y_0)$ is necessary  and sufficient condition
for the element $r\cdot (x_0,y_0)$ of $\Cal J$ to lie
in $\iota(C)$ again.
We emphasize that the function $\psi_n$ is defined over
the curve itself $\iota(C)$ rather than $\Cal J$. This
definition is key of these studies and was, first, given by
Grant \cite{G}.

Cantor showed that
the $\psi_n$-function can be expressed
 in terms of Hankel determinant [C].
However his expression slightly needs a correction on its
factor on $y$ and in \cite{M\^O, Ma2},
we will give a correction along the line of argument of \^Onishi
\cite{\^O2,M\^O}.
(For example, $\psi_2$ of genus two
must be proportional to $y$ from the argument
in [\^O1] but in [C] $\psi_2$ is constant.)
In our new derivation of $\psi$'s \cite{M\^O}, the
Toelpliz determinant is more
natural than the Hankel determinant (see Appendix).
Thus in this article, instead of Hankel determinant,
we use a  Toelpliz determinant,
$$
T_{n}^{(m)}\left( g(s), \frac{d}{ds}\right)
=\left| \matrix
           g^{[m+n-1]} &g^{[m+n-2]} & \cdots
         & g^{[m+1]}& g^{[m]}\\
        g^{[m+n+1]} &g^{[m+n-1]} & \cdots
         & g^{[m+2]}& g^{[m+1]}\\
\vdots&\vdots&\ddots&\vdots&\vdots\\
g^{[m+2n-3]} &g^{[m+2n-4]} & \cdots
        & g^{[m+n-1]}& g ^{[m+n-2]}\\
           g^{[m+2n-1]} &g^{[m+2n-3]} & \cdots
         & g^{[m+n]}& g ^{[m+n-1]}
\endmatrix \right|, \tag 3-5
$$
and $T_{1-n}^{(m)}\left( g(s), \dfrac{d}{ds}\right)\equiv0$
where $m$ and $n$ are positive integers,
$g(s)$ is a function of an argument $s$ and
$$
        g^{[n]}(s):=\frac{1}{n!}\frac{d^n}{d s^n} g(s). \tag 3-6
$$
As we showed in [M\^O], we have an expression of $\psi_n$ as
(see Appendix),
$$
        \psi_n(u) =\left\{\matrix y^{n(n-1)/2}
        \cdot T_{(n-3)/2}^{(4)}( y, \frac{d}{d x}) &
         \text{ for $n$ odd}\\
        y^{n(n-1)/2}\cdot T_{(n-2)/2}^{(3)}( y, \frac{d}{d x})
       & \text{ for $n$ even} \endmatrix \right. . \tag 3-7
$$

Noting $y^2 =f(x)$, $y^{2n-1} d^n y/d x^n$ is a polynomial of $x$
and coprime to $f(x)$ in general.
Hence $y^{n(2m+2n-3)} T^{(m)}_n$, or
$$
\left| \matrix
          y^{2m+2n-3} y^{[m+n-1]} &y^{2m+2n-5} y^{[m+n-2]}
 & \cdots &y^{2m+1}  y^{[m+1]}& y^{2m-1} y^{[m]}\\
        y^{2m+2n-1} y^{[m+n]} & y^{2m+2n-3} y^{[m+n-1]}
& \cdots & y^{2m+3} y^{[m+2]}& y^{2m+1} y^{[m+1]}\\
\vdots&\vdots&\ddots&\vdots&\vdots\\
y^{2m+4n-7} y^{[m+2n-3]} &y^{2m+4n-9} y^{[m+2n-4]} & \cdots
          & y^{2m+2n-3}  y^{[m+n-1]}
& y^{2m+2n-5}  y ^{[m+n-2]}\\
y^{2m+4n-3} y^{[m+2n-2]} &y^{2m+4n-7} y^{[m+2n-3]}
 & \cdots & y^{2m+2n+1}  y^{[m+n+1]}     & y^{2m+2n-3} y ^{[m+n-1]}
\endmatrix \right| \tag 3-8
$$
is an element of $\CC[x]$ and coprime to $y^2$.
Hence $\psi_n(u)$ can be expressed by
$$
        \psi_n =\left\{ \matrix 8 y^3 \alpha_n(x)
               & \text{ for } n=\text{odd }\\
         2 y \alpha_n(x) & \text{ for } n=\text{even }
     \endmatrix \right.
,           \tag 3-9
$$
where $\alpha_n(x)$ is a polynomial of $x$ and coprime of $y$.

Cantor, essentially, showed  that
this $\psi_n$ obeys a recursion relation, which is
an extension of (2-4)  [C, Ma2],
$$
 \psi_2^2 \psi_m\psi_n \psi_{n+m} \psi_{m-n}
=\left| \matrix
 \psi_{m-2}\psi_n & \psi_{m-1}\psi_{n+1} & \psi_m\psi_{n+2}\\
\psi_{m-1}\psi_{n-1} & \psi_m\psi_n & \psi_{m+1}\psi_{n+1}\\
\psi_m\psi_{n-2} &\psi_{m+1}\psi_{n-1}& \psi_{m+2}\psi_n
\endmatrix \right|. \tag 3-10
$$
(3-10) slightly differs from original one in \cite{C}
because original one needs a correction \cite{Ma2, M\^O}.
We emphasize that this relation (3-10)
characterizes the $\psi$ function over a genus two curve
and
 cannot be reduced to some identities
of elliptic functions in general because
(3-10) holds due to the addition relations of
genus two [Ma2].

Noting $\psi_0\equiv 0$, $\psi_1\equiv 0$, $\psi_2 = 2y$
and $\psi_3 = 8y^3$,
we have $n=3$ case of (3-11),
$$
\split
\psi_3 \psi_2^2  \psi_{m+3}\psi_{m-3} \psi_m &-
\psi_3^3 \psi_{m+2}\psi_m \psi_{m-2}
-\psi_5\psi_2^2 \psi_{m+1}\psi_m \psi_{m-1}\\
&+\psi_4\psi_3\psi_2
(\psi_{m-2}\psi_{m+1}^2 + \psi_{m+2}\psi_{m-1}^2 )
=0,
\endsplit
 \tag 3-11
$$
and $n=4$ case,
$$
\split
\psi_4 \psi_2^3\psi_{m+4}\psi_{m-4} \psi_m  &-
\psi_4^3 \psi_{m+2}\psi_m \psi_{m-2}
+\psi_6\psi_4\psi_2 \psi_m^3
-(\psi_5^2\psi_2+\psi_6\psi_3^2)
\psi_{m+1}\psi_m \psi_{m-1}\\
&+\psi_5\psi_4\psi_3
(\psi_{m-2}\psi_{m+1}^2 + \psi_{m+2}\psi_{m-1}^2 )
=0.
\endsplit\tag 3-12
$$
(3-11) and (3-12) are reduced to [C],
$$
\split
\psi_4\psi_2^2\psi_{n+4}\psi_{n-4} &-
        \psi_5\psi_3\psi_2^2\psi_{n+3}\psi_{n-3}
+( \psi_5\psi_3^3-\psi_4^3\psi_2)\psi_{n+2}\psi_{n-2}\\
&+\psi_6\psi_3^2\psi_2\psi_{n+1}\psi_{n-1}
-\psi_6 \psi_4\psi_2^2\psi_{n}\psi_{n}=0.
\endsplit\tag 3-13
$$
This bilinear difference
equation (3-13) is a genus two analog of the bilinear
equation (2-5) and
also characterizes the $\psi$ function over a genus two curve.
We emphasize that from point of view of bilinear
difference equations, (3-13) is a very natural extension
of (2-5) and can not be reduced to a three-term relation
in general.

From here, we will consider modifications of
(3-10)-(3-13) to find
 difference equations which have
form,
$$
        x_{n+m}x_{n-m'} =
          \frac{f_1(x_{n+m-1}, \cdots, x_{n-m'+1})}
               {f_2(x_{n+m-1}, \cdots, x_{n-m'+1})},
\tag 3-14
$$
where $f$'s are polynomials of
$x_{n+m-1}, \cdots, x_{n-m'+1}$ and $m, m'=1,2,3,\cdots$,
by deforming (3-13) as a generalization of (2-6).

By setting $b_n = \psi_{n+2}\psi_{n-2}/\psi_{n}^2$,
(3-13) becomes a sixth order ordinary difference equation,
$$
\split
\psi_4\psi_2^2 b_{n+3}b_{n-3} &-
      \frac{1}{b_{n-2}^2b_{n-1}^3b_{n}^4 b_{n+1}^3b_{n+2}^2}
\Bigr( \psi_5\psi_3\psi_2^2
b_{n-2}b_{n-1}^2b_{n}^3 b_{n+1}^2b_{n+2}\\
&-( \psi_5\psi_3^3-\psi_4^3\psi_2)b_{n-1}b_{n}^2 b_{n+1}
+\psi_6\psi_3^2\psi_2b_{n}
-\psi_6 \psi_4\psi_2^2\Bigr)=0,
\endsplit\tag 3-15
$$
for a general point in $\iota(C)$.
(3-15) can be regarded as a generalization of (2-6).

Further we will consider other difference equations by
dealing with special points.
First we will deal with points $(x,0)$, $y=0$.
For the point, (3-13) becomes
$$
\alpha_{n+4}\alpha_{n-4} -
( \alpha_4^2)\alpha_{n+2}\alpha_{n-2}
+\alpha_6 \alpha_{n}\alpha_{n}=0. \tag 3-16
$$
When we define $c_n = \alpha_{n+2}\alpha_{n-2}/\alpha_{n}^2$,
(3-16) turns out to be the discrete Painlev\'e I (1-1),
$$
        c_{n+2}c_{n-2} = \frac{\alpha_4^2}{c_{n}}-
 \frac{\alpha_6}{c_{n}^2}. \tag 3-17
$$
In other words, we have a hyperelliptic function solution
of the discrete Painlev\'e I [ORGT] (1-1) again,
$$
c_{n}(u) = \frac{\sigma((n+2)u)\sigma((n-2)u)}
            {\sigma(nu)^2 \sigma_2(u)^8} , \tag 3-18
$$
where $u$ is $\iota((x,0))$.

Further using (3-16), we have that
for a point $u \in\iota((x,0))$,
$$
        d_m(u):=\frac{\psi_{m+1}(u)\psi_{m-2}(u)}
{\psi_m(u)\psi_{m-1}(u)}
             = \frac{\sigma((m+1)u ) \sigma((m-2)u )}{
        \sigma(mu)\sigma((m-1)u ) \sigma_2(u)^4}
\tag 3-19
$$
obeys the third order difference equation,
$$
        d_{m+2}d_{m-1} -\frac{ \alpha_5}{d_{m+1} d_m}
                + \alpha_4
          \left( \frac{1}{d_{m+1}}+\frac{1}{d_m}\right)=0.
 \tag 3-20
$$
We note that although $\psi_n$ vanishes for a point $\iota((x,0))$,
$d_m$ has non-trivial values because it can be expressed by
 $$
        d_m(u):=\frac{\alpha_{m+1}(u)\alpha_{m-2}(u)}
{\alpha_m(u)\alpha_{m-1}(u)}.
\tag 3-21
$$
In other words, the third order
difference equation (3-21) is well-defined
and its solution is associated with an algebraic curve.
(3-21) can be also regarded as a generalization of (2-6).

Next we will deal with  a point satisfying $\psi_4=0$.
Then we apparently obtain a forth order
ordinary difference equation,
$$
\psi_5\psi_3\psi_2^2 b_{n+2}b_{n-2} +
      \frac{1}{b_{n-1}^2b_{n}^3 b_{n+1}^2}
\Bigr(
\psi_5\psi_3^3 b_{n-1}b_{n}^2 b_{n+1}
-\psi_6\psi_3^2\psi_2b_{n}\Bigr)=0,
\tag 3-22
$$
at the point. However if $\psi_4$ vanishes at a point $P$
and $P$ is a forth cyclic point,
 it means that $\psi_2\equiv\psi_3\equiv \psi_4\equiv 0$,
$\psi_3\equiv\psi_4\equiv \psi_5\equiv 0$ or
$\psi_4\equiv\psi_5\equiv \psi_6\equiv 0$ from the definition
of $\psi$ function of genus two [C,\^O1].
Thus after fixing the curve (3-1),
we need more precise consideration on (3-22)
whether it has non-trivial meaning or not.

\tvskip
\centerline{\twobf \S 4. Discussion}\tvskip

As mentioned in the introduction, the authors in [HYK] and
[YKTH] have a list of
ordinary third order difference equations.
Our third order difference equation (3-20) also appears in it.
Yahagi shows  [Y] that (3-20) is a special case of (86)
in [HYK],
$$
        x_{m+2}x_{m-1}
        =\frac{a_0 + a_1 x_n +a_1 x_{n+1} +a_3 x_n x_{n+1} }
              {a_3 + b_1 x_n + b_1 x_{n+1} + b_3 x_n x_{n+1} }.
 \tag 4-1
$$
where $a$'s and $b$'s are parameters. (4-1) becomes (3-20)
by letting $ a_3\equiv b_1 \equiv 0$, $a_1\equiv -\alpha_4$
and  $a_0\equiv \alpha_5$.
Though I started this study to find a hyperelliptic function
solutions of the third order difference
equations whose solutions can not be expressed by
elliptic functions, our attempt failed.
As $\psi_n$ at $\iota((x,0))$ becomes the discrete Painlev\'e
equation I, which has the elliptic function solution,
(3-20) is reduced to the ordinary second order
 difference equation [YKTH, Y].

However (3-20) is an equation at the special points
satisfying $y=0$.
We have the  bilinear equation (3-13) as an extension of
(2-5) and (3-10) as that of (2-4) and
(3-10), (3-13), (3-15) (and
(3-22)) cannot be expressed by some
identities of elliptic functions in general.
Even though they cannot be reduced to the third order
difference equation expect (3-20),
it is expected that they might be connected with
higher integrable difference equations.

Further it is also expected that
the recursion relations of $\psi$-functions in more general
algebraic curves, if exist, might contains interesting
several difference equations.
In fact, even though it is not correct by primitive
considerations, a naive extension of
the right hand side of (2-4) and (3-10)
 might be given by
$$
\left| \matrix
\psi_{m-3}\psi_n &  \psi_{m-2}\psi_{n+1} &
       \psi_{m-1}\psi_{n+2} & \psi_m\psi_{n+3}\\
\psi_{m-2}\psi_{n-1} &  \psi_{m-1}\psi_n &
       \psi_{m}\psi_{n+1} & \psi_{m+1}\psi_{n+2}\\
\psi_{m-1}\psi_{n-2} & \psi_{m-1}\psi_{n-1} &
       \psi_{m+1}\psi_n & \psi_{m+2}\psi_{n+1}\\
\psi_{m}\psi_{n-3} & \psi_m\psi_{n-2} &
       \psi_{m+2}\psi_{n-1}& \psi_{m+3}\psi_n
\endmatrix \right|.  \tag 4-2
$$
which looks to contains various third order difference operator.
Thus I believe that it is important to study the recursion
relations of $\psi$ functions over more general algebraic
curves [C, Ma2, M\^O, \^O3, \^O4].

Next we will comment on a relation of our theory to
so-called Sato theory [SN, SS].
As shown in [M\^O],
in the derivation of (3-7)
we encounter the Wronskian.
We will give a rough sketch of the derivation of the proof
in the Appendix.
On the other hand
in the Sato theory, the Wronskian also plays
the important roles \cite{SN, SS}.
In the theory, we encounter a differential equation,
$$
    \frac{\partial W_m }{ \partial t_m} = B_n \cdot
    W_m -W_m \cdot \partial^n,\tag 4-3
$$
where
$W_m = \partial^m + w_1 \partial^{m-1} +\cdots+ w_m$,
$\partial:= \partial/\partial t_1$,
$w_a$ are functions of the Sato coordinate $t=(t_1,t_2,\cdots)$,
and $B_n$ is a certain differential operator [SN].
For the independent solutions $\{f_i\}$
of $W_m(t_i=0,i>1) f(t_1)=0$,
we can define a Wronskian,
$$
        \tau^{(n)}(t) ={\det}_{(n-1)\times(n-1)}
        ( \partial^j f_i(t_1) ).   \tag 4-4
$$
Roughly speaking this Wronskian becomes  $\tau$ function,
which plays central roles in modern soliton theory.
$\tau$ is defined over a Jacobi variety of corresponding
algebraic curves [Mul]. In the theory,
algebraic curve is an auxiliary object after we obtain
the Jacobian.
As the Jacobian is realized as a quotient
complex vector space divided by a discrete lattice,
the $\tau$ function is also related to a discrete equation of
a formula of Jacobi variety $\Cal J$, which is known as
Fay trisecant formula [F].

On the other hand, our theory is of  functions over a curve
 $\iota(C) \subset \Cal J$ based on theory of
 hyperelliptic sigma function \cite{B1-B3}.
As I pointed out in [Ma1],
in Sato theory we consider
the behavior of the differential form of the second kind around
the infinite point, whereas
in Baker's theory, the differential of the first kind, which
is homomorphic all over the curve, plays central roles.
The $\tau$ function can not be explicitly expressed in general
except soliton solutions or elliptic function solutions,
but in Baker's theory
all quantities  concretely can be expressed any points on
a curve \cite{Ma1,BEL}.
The Baker's sigma function theory and $\tau$ function theory are
resemble but  slightly different.

Further in general, functions
 defined over Jacobi variety, {\it i.e.},
$\tau$ and $\sigma$ functions, do not behave well over
 a curve itself;  they vanish or diverge at $\iota(C)$.
Using the properties of $\sigma$ functions,
we can tune the functions and  define $\psi$ functions
over there as in (3-10) [\^O1, \^O2, Ma2].
In our theory, important quantities are defined over a curve itself.
For example, the Wronskian (A-10) appearing in the
derivation of (3-5) in [M\^O] is a  function over a curve
except infinite point.
Thus $\tau$ and $\psi$ functions are also different.
Similarly I think that (3-11) can not be obtained by a ordinary
 modification of Fay
trisecant formula [F]. In other words, I believe that our theory
has an aspect which
one cannot reach by using Sato theory.

\tvskip
{\centerline{\bf{ Acknowledgment}}}
\tvskip

I thank Prof.~Y.~\^Onishi for leading me this beautiful theory
of Baker.
I am grateful to Prof. R. Hirota, Dr. S. Tsujimoto and
Yahagi for helpful comments and sending me their unpublished
article.

\vskip 0.5 cm
\centerline{\twobf \S Appendix  Determinat Expressions of $\psi$-functions}
\vskip 0.5 cm

In this appendix, we will roughly review of
the  dterminant expression of $\psi$-functions
for genus two along the line of the arguments in \cite{M\^O}
in order to show that the $\psi$-function is related to
Wronskian, which is similar to the situation of Sato theory [SN, SS].

As $\psi$-function is a polynomial in $x$ and $y$ over the
curve $C$ (3-1),
the polynomial belongs to an algebra 
$R = \CC[x,y]/( y^2 - f(x) )$.
For the zeros $(x_i, y_i)$ $( i=1, \cdots, n^2 - 1 )$
of $\psi_n(x,y)$, there exists a point $P$ in the curve
$C$,
$n\cdot (x_i,y_i)$ belongs to $\iota(c)$,
$
       n\cdot (x_i,y_i) - n \cdot \infty
        =  P - \infty$ in the sense of Picard group
Pic${}^0(C)$.
In other words, for a local parameter $u_2$ of $\iota(C)
\subset \Cal J$,
$u=(u_1(u_2),u_2)$,
there exist functions $f$, $h \in R$ over the curve
 satisfying,
$$
        f(v_2) ={v_2^n}h(v_2), \tag A-1
$$
or
$$
        \partial_{u_2}^{m} f |_{v_2=0} = 0,
        \quad 0< m < n, \tag A-2
$$
for $v_2=u_2- u_2((x_i,y_i))$, where
$\partial_{u_2}:=(\partial/\partial{u_2})|_{\bar u_2}$.
I have employed the notations $\partial_{u_2}$ as a
partial differential following the notations of
theory of a complex manifold. However it should be
noted that $u_2$ and $v_2$ are local parameter of the curve $C$,
and $u_1$ is a function of $u_2$. Thus we can replace
$\partial_{u_2}$ with the ordinary differential operator $(d/d{u_2})$
as we regard $R$ as a set of functions only of $u_2$ (and $v_2$).
Further we notice that as $\Cal J$ can be identified with Picard group
Pic${}^0(C)$, the addition in this appendix
is defined in $\Cal J$;
 generally we cannot define addition in $\iota(C)$ and
$C$ itself.

The elements in $R$ has a natural order with respect to
the order of divergence at the infinity point,
which is a natural extensions of  the degrees of polynomial.
We regard  $R$ as $\CC$-vector space generated by an ordered set,
$R\equiv< 1, x, x^2, y, x^{3}, xy, x^{4}, x^2 y,
\cdots >$. By introducing new notations,
$$
   \varphi_m =\left\{ \matrix x^s, & \text{ for } & m \le 2\\
         x^p, & \text{ for } &
             p = ( s + 2 )/2,\ m > 2, \ m \text{ is even} \\
 x^q y, & \text{ for } &
       q = ( s - 3)/2,\ m > 2,\ m \text{ is odd}
                                    \endmatrix \right. ,
                                     \tag A-3
$$
$
 R=< 1, \varphi_1, \varphi_2, \cdots.>$.
Let us consider its one-forms,
$$
        d R^{(n)} = < d  \varphi_1,
        d\varphi_2, \cdots, d \varphi_{n-1}>.
         \tag A-4
$$
Each $d \varphi_m$ is expanded by,
$$
        d \varphi_m = \frac{\partial}{\partial u_2} \varphi_m\cdot
        d u_2 +\frac{\partial}{\partial u_2}^2 \varphi_m\cdot
        (d u_2)^{\otimes 2}
        +\frac{\partial^3}{\partial u_2^3} \varphi_m\cdot
        (d u_2)^{\otimes 3}+ \cdots.  \tag A-5
$$
We introduce its truncated expansion,
$$
        d \varphi_m^{(n)} := \frac{\partial}{\partial u_2}
         \varphi_m\cdot
        d u_2 +\frac{\partial^2}{\partial u_2^2} \varphi_m\cdot
        (d u_2)^{\otimes 2} + \cdots
        +\frac{\partial^{n-1}}{\partial u_2^{n-1}} \varphi_m\cdot
        (d u_2)^{\otimes (n-1)}.  \tag A-6
$$
and a $(n-1)$-dimensional $\CC$-linear space,
$$
      V^{(n)} = <d u_2 ,(d u_2)^{\otimes 2},
       \cdots,(d u_2)^{\otimes( n-1)}>.  \tag A-7
$$

Let us consider the $\CC$-linear map at each point $u$,
$$
        \xi^{(n)} : V^{(n)} \to d R^{(n)}.  \tag A-8
$$
At the zero of $\psi_n$, $(x_i,y_i)$, (A-2) implies
that there exists
a polynomial $\phi^{(n)} := \sum_{i=s}^n a_m \varphi_m$
($a_m \in \CC$),
$$
        \frac{\partial^m}{\partial u_2^m} \phi^{(n)}
        |_{u_2=u_2(x_i,y_i)} = 0,
        \quad 0< m < n. \tag A-9
$$
It means that
$ d \phi^{(n)} := \sum_{i=m}^n a_m d \varphi_m^{(n)} $
vanishes at $u_2=u_2(x_i,y_i)$. In other words, the rank of
the map $\xi^{(n)} $
is smaller than $n-1$ and the Jacobian of $\xi^{(n)} $,
$$
        J^{(n)}(u) ={\det}_{(n-1)\times(n-1)}
        \left( \frac{\partial^j}{\partial u_2^j}\varphi_i(u)
 \right)   \tag A-10
$$
vanishes. This $J^{(n)}(u)$ can be
regarded as the Wronskian
and as a coefficient of
a natural form $(d u_2)^{\otimes k}$'s,
$W^{(n)}(u) := J^{(n)}(u)\cdot (d u_2)^{\otimes n(n-1)/2}
$.

Thought $J^{(n)}(u)$ is proportional to  $\psi_n$,
$ J^{(n)}(u)$ is
 not polynomial in $x,y$ in general.
Thus we will modify it using
a formula $\dfrac{\partial}{\partial u_2}
 = \dfrac{y}{x^{}} \dfrac{\partial}{\partial x}$ in $\Cal J$
and a relation over $\iota(C)$,
$$
{\det}_{(n-1)\times(n-1)} \left( \frac{\partial^j}{\partial 
u_2^j}\varphi^{(i)}
\right)
 =\left(\frac{y}{x^{}}\right)^{ n(n-1)/2}
 {\det}_{(n-1)\times(n-1)} \left( \frac{\partial^j}{\partial 
x^j}\varphi^{(i)}
  \right)
.
     \tag A-11
$$
We note that
as  $R$ is a set of functions of $x$ (not of $\bar x$), we can
also replace
$\partial^j/\partial x^j$ with an ordinary differential operator
$d^j/d x^j$.
It turns out that $ x^{(g-1) n(n-1)/2} J^{(n)}(u)$ is a polynomial
 of $(x,y)$.
Accordingly we define
$$
        \psi_n(u) = \gamma x^{ n(n-1)/2} J^{(n)}(u),  \tag A-12
$$
where $\gamma$ is a constant factor, which will be determined later.

Here we permute the generator of $R$
 in terms of a permutation group $\frak S$,
$$
\split
d \tilde R^{(n)}&=\frak S d R^{(n)} = < d x, d (x^2),
\cdots, d (x^p),
        d y, d(xy), \cdots, d (x^q y)>\\
            &=: < d  \tilde\varphi_1,
        d\tilde\varphi_2, \cdots, d \tilde\varphi_{n-1}>.
\endsplit
         \tag A-13
$$
where $p\in \NN$ and $q \in \ZZ_{\ge 0}$;
$$
       p=
\left\{ \matrix n-1, & \text{ for } & n \le 3,\\
         \left[\frac{n+1}{2}\right], & \text{ for } &
              n > 3,     \endmatrix \right.
$$
$$
       q=
\left\{\matrix \text{not defined}, & \text{ for } & n \le 3,\\
         \left[\frac{n-4}{2}\right], & \text{ for } &
              n > 3.   \endmatrix \right. \tag A-14
$$

$$
\centerline{
\vbox{
        \baselineskip =10pt
        \tabskip = 1em
        \halign{&\hfil#\hfil \cr
        \multispan7 \hfil Table 1: $g=2$ \hfil \cr
        \noalign{\smallskip}
        \noalign{\hrule height0.8pt}
        \noalign{\smallskip}
& $r$ &\strut\vrule& 1  & 2  & 3 & 4 & 5 & 6 & 7 & 8 & 9  \cr
\noalign{\smallskip}
\noalign{\hrule height0.3pt}
\noalign{\smallskip}
& $p$ & \strut\vrule&  0 & 1 & 2 & 2 & 3 & 3 & 4 & 4 & 5 \cr
& $q$ & \strut\vrule&  - & - & - & 0 & 0 & 1 & 1 & 2 & 2 \cr
& $p-q+1$ & \strut\vrule&  - & - & - & 3 & 4 & 3 & 4 & 3 & 4 \cr
\noalign{\smallskip}
        \noalign{\hrule height0.8pt}
}
}
}
$$

The quantity
$ {\det}_{(n-1)\times(n-1)} ( \partial_x^j\tilde\varphi^{(i)} )
$
is computed by
$$
\split
\tilde J^{(n)}&=
\left| \matrix
1 & 2 x  & \cdots & p x^{p-1} &
               \dot y &\dot{( xy)} & \cdots &\dot{(x^{q} y )}\\
\  & 2 &  \cdots & p(p-1) x^{p-2} &
               \ddot y &\ddot{( xy) }& \cdots &\ddot{(x^{q} y )}\\
\  &\   & \ddots &\vdots&\vdots&\vdots&\cdots&\vdots\\
\    & \ & \   & p! &
           y^{(p)} &( xy)^{(p)} & \cdots &(x^{q} y )^{(p)}\\
\    & \ & \  &\ &
           y^{(p+1)} &( xy)^{(p+1)} & \cdots &(x^{q} y )^{(p+1)}\\
\   &\  & \   &\  &\vdots&\vdots&\cdots&\vdots\\
\    & \ & \   &\ &
           y^{(n-1)} &( xy)^{(n-1)} & \cdots &(x^{q} y )^{(n-1)}
\endmatrix \right|\\
 &=
\left| \matrix
1 & \ & \ & \ & \     & \  &
              \  &\  & \     &\ \\
\  & 2 & \ & \ & \     & \  &
                \  &\  & \     &\ \\
\   &\  & 3! &\   & \     &\    &\    &\    &\    &\    \\
\   &\   &\ & \ddots  & \     &\    &\    &\    &\    &\    \\
\  & \ & \  & \ & \     & p! &
           \  &\   & \     &\  \\
\  & \ & \  & \ & \     &\ &
           y^{(p+1)} & {}_{p+1} P_1 y^{(p+2)} & \cdots
         & {}_{p+1} P_{q-1} y^{(p-q)} &{}_{p+1} P_{q} y^{(p-q+1)}\\
\    &\  & \ &\  & \  &\  &\vdots&\vdots&\cdots&\vdots&\vdots\\
\  & \ & \  & \ & \     &\ &
           y^{(n-2)} &{}_{n-2} P_1 y^{(n-3)} &
       \cdots &{}_{n-2} P_{q-1}y ^{(p+1)} & {}_{n-2} P_{q}y ^{(p)}\\
\  & \  & \ & \ & \     &\ &
           y^{(n-1)} &{}_{n-1} P_1 y^{(n-2)}
       & \cdots &{}_{n-2} P_{q-1} y ^{(p)}&{}_{n-2} P_{q} y ^{(p+1)}
\endmatrix \right|\\
 &=1! 2! \cdots p! (p+1)! (p+2)! \cdots (n-1)! \cdot
 T_{q+1}^{(p-q+1)}(y, \frac{d }{dx}),
\endsplit \tag A-15
$$
where ${}_n P_m:=n!/(n-m)!$.
Let $\gamma=\det (\frak S)/ 1! 2! \cdots (n-1)!$ in (A-12)
and then we have
$$
        \psi_n(u) =  y^{n(n-1)/2}
        T_{q+1}^{(n-1)}(y, \frac{d }{dx}). \tag A-16
$$
This is reduce to (3-7) and
 $\psi_n$ can be regarded as a Wronskian.

\enddocument

\enddocument